# On the paper "Role of potentials in the Aharonov-Bohm effect"

## A. M. Stewart

*Emeritus Faculty, The Australian National University, Canberra, ACT 0200, Australia.*

*http://www.anu.edu.au/emeritus/members/pages/a_stewart/*

**Abstract:** When the magnetic vector potential is expressed in terms of the magnetic field it, is found to be explicitly non-local in space. This gives support to the conclusions of Aharonov *et al.* in a recent comment, that the Aharonov-Bohm effect may be interpreted as being either due to a local gauge potential or else due to non-local gauge-invariant fields but *not* due to local gauge-invariant fields.

## I. INTRODUCTION

The role of non-locality in the Aharonov-Bohm (AB) effect [1-5] continues to attract attention. The issue in question is why the interference pattern of an electron diffraction experiment should be affected by magnetic fields when the electrons are constrained to pass through a region of space in which the electromagnetic gauge fields **E** or **B** are zero but in which their scalar and vector electromagnetic gauge potentials V and **A** are non-zero. Recently Vaidman [6] argued that that when the source of the electromagnetic potential is treated in the framework of quantum theory, the Aharonov-Bohm effect can be explained without the notion of potentials. Aharonov, Cohen, and Rohrlich [7] challenged this view but their arguments were rejected by Vaidman [8]. In this paper we argue that, because the potentials are found to be *explicitly* non-local in space, the conclusions of Aharonov *et al.* that the AB effect may be interpreted as being *either* due to a local gauge potential *or else* due to non-local gauge-invariant fields but *not* due to local gauge-invariant fields is correct. We consider the static magnetic AB effect.

## II. NON-LOCALITY

The standard relation between gauge-invariant magnetic gauge field **B** and the gauge-variant vector gauge potential **A** is

$$\mathbf{B}(\mathbf{r},t) = \nabla \times \mathbf{A}(\mathbf{r},t) \quad , \quad (1)$$

where $\nabla$ is the gradient operator with respect to **r**. The magnetic vector potential **A** can, as can any physically realistic 3-vector [9-12], be expressed as a sum of the gradient of a scalar field and the curl of a vector field, a decomposition that is unique [13, 14]. The gradient part may be ignored here; it contributes nothing to the AB phase because it is integrated around a closed path. It also contributes nothing to **B** because the curl of a gradient is zero.

The curl part, when expressed in terms of the **B** field that it gives rise to, is [12, 15, 16]

$$\mathbf{A}(\mathbf{r},t) = \nabla \times \int d^3r' \frac{\mathbf{B}(\mathbf{r}',t)}{4\pi |\mathbf{r} - \mathbf{r}'|} \quad . \tag{2}$$

This expression may be obtained by appealing to the Helmholtz theorem of vector decomposition [12, 15, 17] or by taking its curl and finding that the result is $\mathbf{B}(\mathbf{r},t)$ [16]. In either method, partial integration is used which requires certain surface integrals to vanish at infinity. It is found that for physically realizable electromagnetic fields these surface integrals do indeed vanish [5, 11]. An analogous expression has also been found for the scalar potential [5]

$$V(\mathbf{r},t) = \nabla \cdot \int d^3r' \frac{\mathbf{E}(\mathbf{r}',t)}{4\pi |\mathbf{r} - \mathbf{r}'|} \quad . \tag{3}$$

Both of these potentials are in the Coulomb gauge ($\nabla \cdot \mathbf{A} = 0$), which has the interesting feature of being a minimal gauge in the sense that the integral of $\mathbf{A}^2$ over all of 3-space is a minimum for this gauge [18, 19]. The potentials (2) and (3) may be transformed to any other gauge by making a gauge transformation [4, 5, 20, 21] but this just adds a gradient term that is irrelevant for the AB effect. The two potentials given in (2) and (3) encode their respective fields at every point in space at the same time. The instantaneous nature of these potentials was queried by Dmitriyev [22] but it was pointed out [16] that because the potentials and the fields propagate at the same speed (of light) from their sources it is no surprise that one can be expressed in terms of the other at the same time.

The potentials (2) and (3) are explicitly non-local in space: the field $\mathbf{B}$ at $\mathbf{r}'$ produces a field $\mathbf{A}$ at $\mathbf{r}$ at the same time. This non-local feature was anticipated by Feynman [23], and demonstrates how, within the realm of classical electrodynamics, it is possible for the potentials to be non-zero in regions where the fields are zero. We therefore agree with the conclusions of Aharonov *et al.* [7] that the AB effect may be interpreted as being *either* due to the gauge potential $\mathbf{A}(\mathbf{r},t)$ that is *local* or else due to gauge-invariant fields $\mathbf{B}(\mathbf{r}',t)$ that are *non-local*, as in the right-hand side of (2). The AB effect cannot be interpreted in terms of gauge-invariant fields that act locally. Although we find that the non-locality is manifested within the framework of classical electrodynamics, the AB effect still relies on quantum mechanics because matter-wave interference is an essentially quantum mechanical effect.

Vaidman [6] has expressed the hope that a general formalism of quantum mechanics based on local fields will be developed. Such a formalism was introduced by DeWitt [24] but was found to be non-local [25]. We argue that it is not possible to express the quantum mechanics of particles interacting with gauge fields in terms of the gauge field at the local position of the particle alone. This is because Hamiltonian quantum mechanics, which is used to describe the AB effect, does, as its name suggests, measure the *energies* of particles in the fields. The energy of a particle in a gauge field is a non-local quantity. For the simplest example, consider the electrostatic energy $V(\mathbf{r})$ of a charge $q$ at $\mathbf{r}$ in an electric field $\mathbf{E}(\mathbf{x})$, which is given by the line integral of -$q\mathbf{E}(\mathbf{x}) \cdot d\mathbf{x}$ from infinity to the position of the charge $\mathbf{r}$, a non-local quantity as it involves $\mathbf{E}$ along the whole path and not just $\mathbf{E}$ at $\mathbf{r}$. There presently is and is unlikely to be no formulation of quantum mechanics that does not involve either local gauge potentials or non-local gauge fields, so attempts to explain the AB effect or any other quantum mechanical effect on the basis of local gauge fields are likely to be unproductive.

APPENDIX A

It is desirable to show that that (2) can indeed describe the vector potential produced by a flux cylinder of finite radius $R$. It has been shown previously [5, 15, 26] that, for a straight infinitely long and thin tube of flux, equation (3) reproduces the Stokes law result $\delta A = \delta \Phi /(2\pi d)$, where $\delta \Phi$ is the flux tube and $d$ is the perpendicular distance between the flux tube and the point at which **A** is evaluated (Fig. 1). Cylindrical coordinates $\{\rho, \theta, z\}$ are used and **A** lies perpendicular to $d$ in the $z$ = constant plane. We need to integrate this result over the cross-section at $z$ of an infinitely long cylinder of finite radius $R$ of flux $\Phi$, containing a uniform magnetic field $B$ so that $\Phi = B\pi R^2$. We evaluate the vector potential at $\{\rho, \theta, z\}$ with $\theta = 0$ due to a flux element $\delta \Phi = B\rho' d\rho' d\theta'$ at $\{\rho', \theta', z\}$ where $0 < \rho' < R$.

From Fig. 1 the angle $\alpha$ between the vector potential due to $\delta \Phi$ and the $\theta = \pi/2$ direction is given by $\sin\alpha = \rho'\sin\theta'/d$. The projection of this along the $\theta = 0$ direction, when integrated over $\theta'$, is zero because the integrand is an odd function of $\theta'$. The projection along the $\theta = \pi/2$ direction comes to

$$\rho' d\rho' \int_{-\pi}^{\pi} d\theta' \cos\alpha \frac{d\Phi}{2\pi d} = \rho' d\rho' \frac{\Phi[1+\text{Sign}(\rho-\rho')]}{2\pi\rho R^2} \qquad (4)$$

When this is integrated over $\rho'$, the Stokes result is obtained: $A = \Phi/(2\pi\rho)$ for $R < \rho$ and $A = \Phi\rho/(2\pi R^2)$ for $0 < \rho < R$, thereby confirming that (2) gives the accepted result for flux cylinders of finite radius. In addition, (2) has the potential to calculate potentials $\mathbf{A}\{\rho, \theta, z\}$ for flux densities that are not uniform.

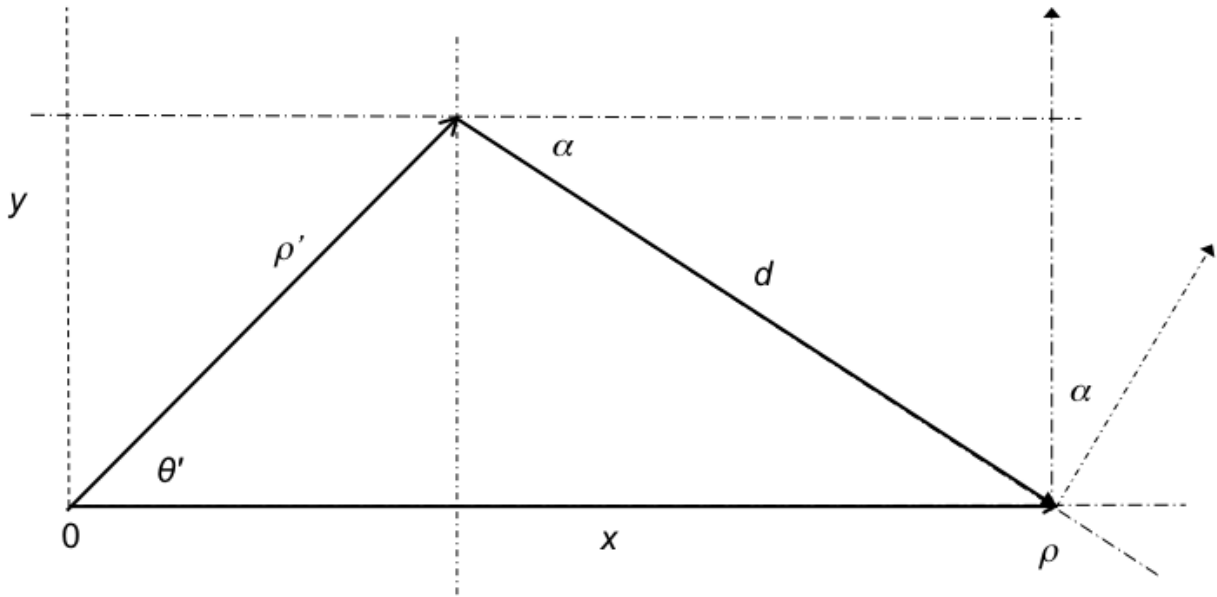

FIG. 1. The figure shows the geometrical relation of an infinitesimal flux tube at $\{\rho', \theta', z\}$ to the vector potential that it contributes to at $\{\rho, \theta = 0, z\}$. Cylindrical coordinates are used.

## APPENDIX B

The validity of (2) has been demonstrated in [15,16]. Equation (3) is shown to be valid by taking its gradient

$$\nabla V(\mathbf{r},t) = \int \frac{d^3 r'}{4\pi} \nabla[\nabla \cdot \frac{\mathbf{E}(\mathbf{r}',t)}{|\mathbf{r}-\mathbf{r}'|}] \quad . \quad (5)$$

to give

$$\nabla V(\mathbf{r},t) = \int \frac{d^3 r'}{4\pi} \nabla[\mathbf{E}(\mathbf{r}',t) \cdot \nabla \frac{1}{|\mathbf{r}-\mathbf{r}'|}] \quad , \quad (6)$$

or

$$\nabla V(\mathbf{r},t) = \int \frac{d^3 r'}{4\pi} [\mathbf{E}(\mathbf{r}',t) \cdot \nabla] \nabla \frac{1}{|\mathbf{r}-\mathbf{r}'|} \quad . \quad (7)$$

Consider the vector identity

$$\nabla \times [\mathbf{E}(\mathbf{r}',t) \times \nabla \frac{1}{|\mathbf{r}-\mathbf{r}'|}] = \mathbf{E}(\mathbf{r}',t) \nabla^2 \frac{1}{|\mathbf{r}-\mathbf{r}'|} - [\mathbf{E}(\mathbf{r}',t) \cdot \nabla] \nabla \frac{1}{|\mathbf{r}-\mathbf{r}'|} \quad , \quad (8)$$

which leads to

$$\nabla V(\mathbf{r},t) = -\mathbf{E}(\mathbf{r},t) - \nabla \times \int \frac{d^3 r'}{4\pi} [\mathbf{E}(\mathbf{r}',t) \times \nabla \frac{1}{|\mathbf{r}-\mathbf{r}'|}] \quad . \quad (9)$$

Next, take the time derivative of (2) and use a homogeneous Maxwell equation to give

$$\frac{\partial \mathbf{A}(\mathbf{r},t)}{\partial t} = -\nabla \times \int \frac{d^3 r'}{4\pi} \frac{[\nabla' \times \mathbf{E}(\mathbf{r}',t)]}{|\mathbf{r}-\mathbf{r}'|} \quad , \quad (10)$$

where $\nabla'$ is the gradient operator with respect to **r'**. Consider the vector identity

$$\nabla' \times [\frac{\mathbf{E}(\mathbf{r}',t)}{|\mathbf{r}-\mathbf{r}'|}] = \mathbf{E}(\mathbf{r}',t) \times \nabla \frac{1}{|\mathbf{r}-\mathbf{r}'|} + \frac{\nabla' \times \mathbf{E}(\mathbf{r}',t)}{|\mathbf{r}-\mathbf{r}'|} \quad . \quad (11)$$

which, provided that the volume integral over **r'** associated with the left hand side of (11) vanishes, gives

$$\frac{\partial \mathbf{A}(\mathbf{r},t)}{\partial t} = \nabla \times \int \frac{d^3 r'}{4\pi} \mathbf{E}(\mathbf{r}',t) \times \nabla \frac{1}{|\mathbf{r}-\mathbf{r}'|} \quad (12)$$

Hence, from (9)

$$\mathbf{E}(\mathbf{r},t) = -\nabla V(\mathbf{r},t) - \frac{\partial \mathbf{A}(\mathbf{r},t)}{\partial t} \quad , \quad (13)$$

as required. Finally, the volume integral of the left-hand side of (11) is

$$-\nabla \times \int \frac{d^3 r'}{4\pi} \nabla' \times [\frac{\mathbf{E}(\mathbf{r}',t)}{|\mathbf{r}-\mathbf{r}'|}] \qquad , \qquad (14)$$

This is transformed into a surface integral at $r' \to \infty$,

$$-\nabla \times \int \frac{d^2 \Omega'}{4\pi} \hat{\mathbf{r}}' \times [\frac{\mathbf{E}(\mathbf{r}',t)}{|\mathbf{r}-\mathbf{r}'|}] r'^2 \qquad , \qquad (15)$$

where $\Omega$ is the solid angle. The $\nabla$ at the front causes the $1/|\mathbf{r}-\mathbf{r}'|$ term to have a $1/|\mathbf{r}-\mathbf{r}'|^2$ dependence, and because radiation fields, the longest-range electromagnetic fields, go as $(\sin kr')/r'$, where $k$ is the wave vector of the radiation [11], the surface integral vanishes as required.